\documentclass[aps,prl,
twocolumn,showpacs,amssymb]{revtex4}
\usepackage{t1enc}
\usepackage[latin1]{inputenc}
\usepackage[english]{babel}
\usepackage{graphics}
\usepackage{graphicx}
\usepackage{amssymb}
\usepackage{dcolumn}
\usepackage{bm}
\usepackage{color}

\begin{document}

\renewcommand{\figurename}{Fig.}

\title{Orbital character variation of the Fermi surface and doping dependent changes of the dimensionality 
in BaFe$_{2-x}$Co$_x$As$_2$ from  angle-resolved photoemission spectroscopy}
\author{S.\ Thirupathaiah,$^1$ S.\ de Jong,$^3$ R.\ Ovsyannikov,$^1$ H.A.\ D\"urr,$^1$ A.\ Varykhalov,$^1$ R.\ Follath,$^1$ Y.\ Huang,$^3$ R.\ Huisman,$^3$ M.S.\ Golden,$^3$ Yu-Zhong Zhang,$^4$ H.O.\ Jeschke,$^4$ R.\ Valent\'{\i},$^4$ A.\ Erb,$^5$ A.\ Gloskovskii,$^6$ and J.\ Fink,$^{1,2}$}
\affiliation{
$^1$Helmholtz-Zentrum Berlin, Albert-Einstein-Strasse 15, 12489 Berlin, Germany\\
$^2$Leibniz-Institute for Solid State and Materials Research Dresden, P.O.Box 270116, D-01171 Dresden, Germany\\
$^3$Van der Waals-Zeeman Institute, University of Amsterdam, NL-1018XE Amsterdam, The Netherlands\\
$^4$Inst. für Theor. Physik, Goethe-Universität, Max-von-Laue-Straße 1, 60438 Frankfurt, Germany\\
$^5$Walther-Meissner-Institut, Walther-Meißner Strasse 8, 85748 Garching, Germany\\
$^6$Institut für Anorganische Chemie und Analytische Chemie, Johannes Gutenberg-Universität, 55099 Mainz, Germany}

\date{\today}

\begin{abstract}
From a combination of high resolution angle-resolved photoemission spectroscopy and density functional 
calculations, we show that BaFe$_{2}$As$_2$ possesses essentially two-dimensional electronic states, 
with a strong change of orbital character of two of the $\Gamma$-centered Fermi surfaces as a function of $k_z$.
Upon Co doping, the electronic states in the vicinity of the Fermi level take on increasingly 
three-dimensional character. Both the orbital variation with $k_z$ and the more  three-dimensional 
nature of the doped compounds have important consequences for the nesting conditions and thus possibly also for 
the appearance of antiferromagnetic and superconducting phases.
\end{abstract}
\maketitle

Since the discovery of high $T_c$ superconductivity in Fe-pnictides \cite{Kamihara:2008},
many experiments have been carried out to reveal
the physical and electronic properties of these materials \cite{Rotter:2008,
Wang:2009,Yuan:2009,Tanatar:2009}. 
The parent compounds of Fe-pnictide superconductors
are antiferromagnetic (AFM) metals. Both electron and
hole doping suppresses the AFM order and leads to a superconducting phase.
The AFM ordering is supposed to occur by nesting of hole pockets 
at the center of the Brillouin zone (BZ)
and electron pockets at the zone corner. Nesting may be also important 
for the pairing mechanism in these
compounds~\cite{Mazin:2008} although there are alternative scenarios based on the high polarizability of the As ions~\cite{Sawatzky:2009}. The nesting scenario could 
explain why in the SmFeAsO-based 
superconductors \cite{Liu:2008}, predicted to have an almost 
two-dimensional electronic structure \cite{Singh:2008,Ma:2008}, higher superconducting transition 
temperatures $T_c$ are observed than
in BaFe$_2$As$_2$-based systems \cite{Rotter:2008} which are predicted to have a more three-dimensional 
electronic structure \cite{Singh:2008-1}. 
In general, reduction of the dimensionality
increases the number of states that could be considered to be well nested. Furthermore, we point 
out that the orbital character of the states at the Fermi level $E_F$ is very important for the nesting conditions
as the interband transitions which determine the electronic susceptibility, as calculated
by the Lindhard function, are (in weak coupling scenarios) by far strongest when the two Fermi surfaces 
have the same orbital character \cite{Graser:2009}. 
The admixture of three-dimensionality, 
arising from interlayer coupling, makes the materials 
potentially more useful in devices and other applications.
Thus the dimensionality of the electronic structure, i.e., the $k_z$ dispersion of the electronic states
is of great importance for the understanding and application of these new superconductors.

Although angle-resolved photoemission spectroscopy (ARPES) is an ideal tool to study the dispersion
of bands parallel and perpendicular to the FeAs layers there exist only a few experimental studies of these 
issues\cite{CLiu:2009,Vilmercati:2009,Kondo:2009}.
In this letter, we report a systematic study of the dimensionality of the electronic structure of 
BaFe$_{2-x}$Co$_x$As$_2$ ($x$= 0 to 0.4) using
polarization dependent ARPES, uncovering two new factors which are of great significance for 
the nesting of the Fermi surfaces of these systems.
Firstly we show that the Co doping of BaFe$_2$As$_2$ strongly increases
the three-dimensionality of the electronic structure. Secondly, we also detect an important change of the orbital
character of the electronic states at the Fermi level when changing the 
wave vector perpendicular to the layers. Our
results are in qualitative agreement with density functional theory (DFT) calculations.

Single crystals of BaFe$_{2-x}$Co$_x$As$_2$, were grown
in Amsterdam using a self-flux method. Another set of single crystals 
of BaFe$_2$As$_2$ were grown in Garching using Sn-flux. Characterizing studies on 
Amsterdam samples have been reported elsewhere ~\cite{Massee}. The ARPES measurements
were carried out at the BESSY II synchrotron radiation facility using the UE112-PGM2a beam line, 
equipped with a SCIENTA SES 100 analyzer. The total energy resolution
was 25 meV while the angular resolution was 0.2$^\circ$ along
the slit of analyzer and 0.3$^\circ$ perpendicular to it. 
All the samples were cleaved {\it in situ} at a temperature of less than 50 K. Further experimental 
details have been published previously \cite{Fink}. Due to matrix element effects, measurements performed with polarized photons can detect different Fe 3$d$ states depending on the polarization and sample alignment (i.e, $\Gamma$-X or $\Gamma$-M) (see Table I).
In the following we will not discuss the $xy$ states as they are well below the Fermi energy $E_F$.
\begin{table}[t]
 \caption{Sample alignment and Fe 3$d$ orbitals which can be detected with s and p-polarized photons.}
	\centering
		\begin{tabular}{c c c c c c}
			\hline\hline
			Alignment & Fe $3d_{x^2-y^2}$ & Fe $3d_{z^2}$ & Fe $3d_{xz}$ & Fe $3d_{yz}$ & Fe $3d_{xy}$ \\
			\hline \\
			$\Gamma$-M (p-pol) & yes & yes & yes & no & no\\
			$\Gamma$-M (s-pol) & no & no & no & yes & yes\\
			$\Gamma$-X (p-pol) & no & yes & yes & yes & yes\\
			$\Gamma$-X (s-pol) & yes & no & yes & yes & no\\
	\noalign{\smallskip}\hline\hline
		\end{tabular}
\end{table}

DFT calculations have been performed on BaFe$_2$As$_2$ 
and BaFe$_{1.8}$Co$_{0.2}$As$_2$, 
using the Perdew-Burke-Ernzerhof generalized gradient approximation (see Fig.~1). 
For BaFe$_2$As$_2$ experimental structural data were taken from \cite{Huang}.
For BaFe$_{2-x}$Co$_x$As$_2$ we use
Car-Parrinello molecular dynamics based on 
projector augmented wave basis to fully optimize the lattice 
structure within the virtual 
crystal approximation. High symmetry points of the BZ are denoted 
by $\Gamma$ = (0, 0, 0), Z = (0, 0, 1), 
X = (1/2, 1/2, 0), and K = (1/2, 1/2, 1) in the units (2$\pi/a$, 2$\pi/a$, 2$\pi/c$), where
$a$ and $c$ are the tetragonal lattice constants of BaFe$_2$As$_2$ along the $x$ and $z$ axis, respectively. 
In Fig.~1(a) and (b), the orbital character of the bands is shown in color.
\begin{figure}[t]
	\centering
		\includegraphics[width=8 cm]{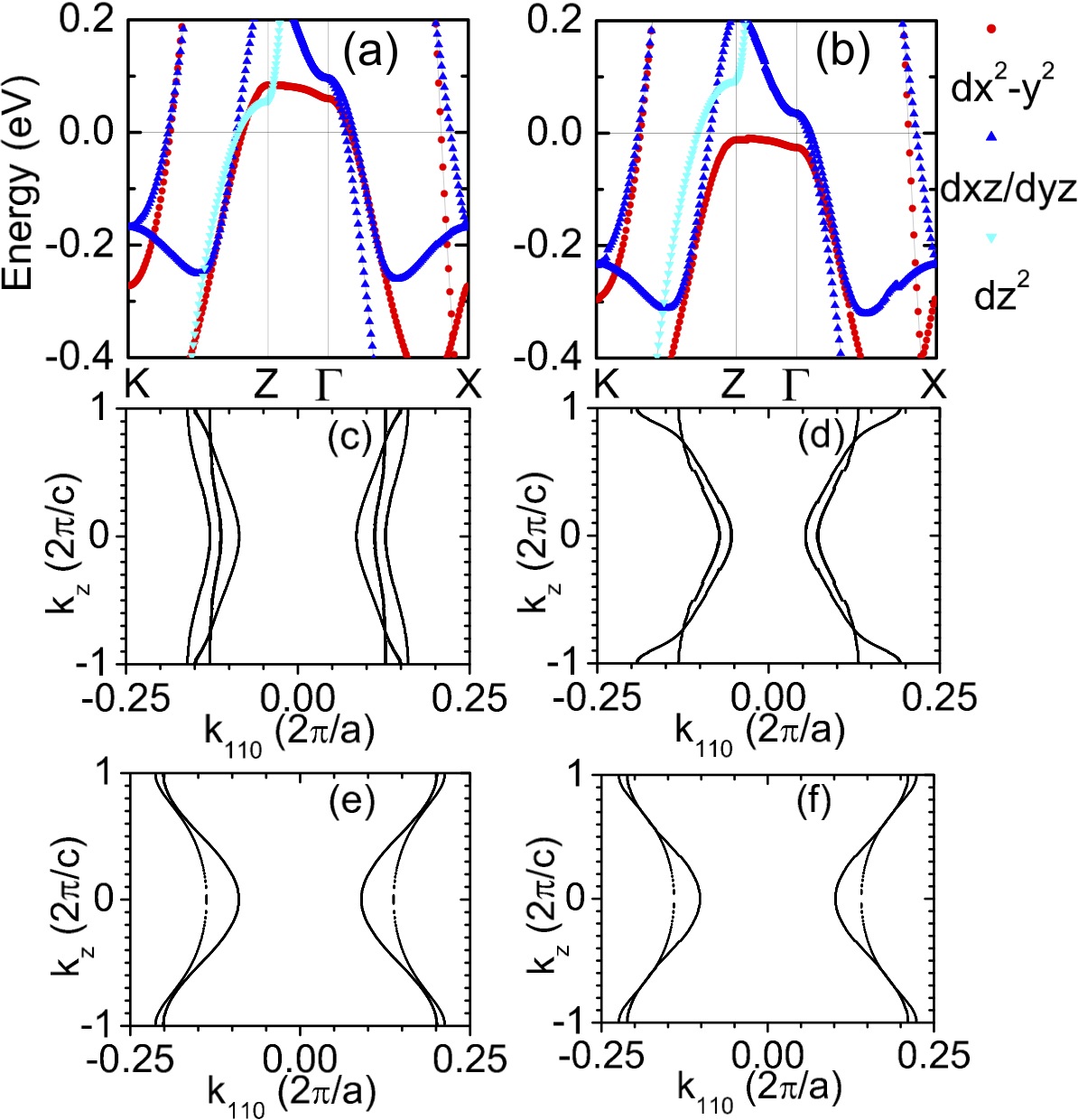}
			\caption{ (color online) Bands and their orbital character (a) for  BaFe$_2$As$_2$ and (b) for 
			 BaFe$_{1.8}$Co$_{0.2}$As$_2$. (c) $k_z$ dispersion around $\Gamma$ for BaFe$_2$As$_2$ and (d) for 
			  BaFe$_{1.8}$Co$_{0.2}$As$_2$. (e) $k_z$ dispersion around X  
			  for BaFe$_2$As$_2$ and (f) for 
			   BaFe$_{1.8}$Co$_{0.2}$As$_2$.}
	\label{fig:2DFT_cal}
\end{figure}

ARPES measurements on undoped BaFe$_2$As$_2$ are displayed in Fig.~2. The measurements were 
performed using the photon energies, h$\nu$=75 eV and h$\nu$=57 eV, corresponding to 
$k_z\approx$0 ($\Gamma$ point) and $k_z\approx$1 (Z point), respectively. 
The $k_z$ values are calculated using an inner potential of 15 eV \cite{Vilmercati:2009}. 
s-polarized photons were used for recording the data along the $\Gamma$-X direction. 
Figure~2(b)-(e) depicts the h$\nu$=75 eV data. 
\begin{figure*}
	\centering
		\includegraphics[width=0.9\textwidth]{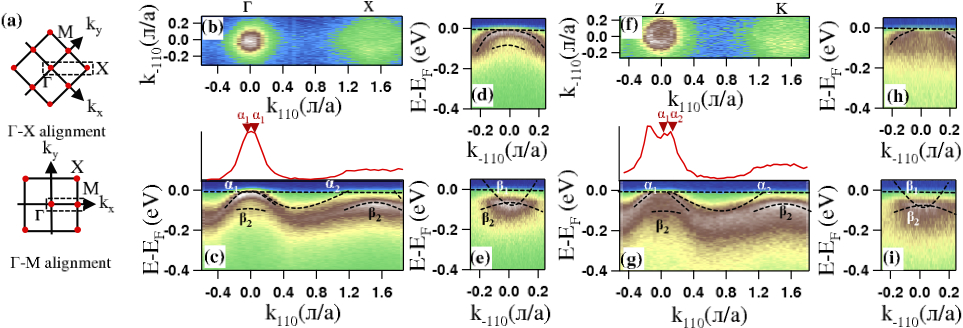}
			\caption{(color online) ARPES data on undoped BaFe$_2$As$_2$ using s-polarized photons measured 
			along $\Gamma$-X. (a) Cartoon of sample alignment and $k$ space covered.  (b)-(e) 
			h$\nu$=75 eV, corresponding to $k_z\approx$0 ($\Gamma$). 
			(b) Fermi surface map, (c) energy distribution map 
			(EDM) taken along the $k_{110}$ direction, (d) EDM along $k_{-110}$ around the 
			$\Gamma$ point and (e) around the X point. (f)-(i) Analogous data but taken 
			with h$\nu$ = 57 eV, corresponding to $k_z\approx$1 (Z). The red curves in panel (c) and (g) 		
			represent momentum distribution curves at $E_F$.}
	\label{fig.1}
\end{figure*} 
We observe a hole pocket at $\Gamma$ and an electron pocket at X in the Fermi 
surface map. Figure~2(c) shows the energy distribution map taken along the $k_{110}$ 
direction. We resolved two bands ($\alpha_1$ and $\alpha_2$) at $\Gamma$. Only 
$\alpha_1$ crosses $E_F$ while $\alpha_2$ is not visible
for binding energies less than 20 meV. A Fermi vector $k_F = 0.07 \pm 0.01 $ {\AA}$^{-1}$ is calculated 
for $\alpha_1$ by fitting the momentum distribution curve (MDC) taken over the integration range 
of a 20 meV window with respect to $E_F$. Using the polarization
dependent selection rules (see Table I), with s-polarized photons 
along the $\Gamma$-X direction, the $\alpha_1$ band can be attributed to ${x^2-y^2}$ 
and the $\alpha_2$ band is related to ${xz}$ and ${yz}$ states. 
Around $\Gamma$ a weak spectral feature is observed near 100 meV below $E_F$. 
This feature is related to a back folding and hybridisation of bands near $\Gamma$ and X, due to 
the AFM order in the As-Fe-As block.
Around the zone corner X we observe two bands along $k_{110}$ ($\alpha_2$ and $\beta_2$) [Fig.~2(c)],
and an additional hole 
pocket along $k_{-110}$ ($\beta_1$) [Fig.~2(e)]. Maintaining the same geometry, but switching to h$\nu$=57 eV 
($k_z\approx$1) we can identify [Fig.~2(f)] two hole pockets around $ \Gamma$ and 
one electron pocket around X. The energy distribution map [Fig.~2(g)] shows two bands crossing $E_F$ 
at the zone center ($\alpha_1$ and $\alpha_2$). 
Since there is no clear separation between these bands 
we give the average Fermi vector $k_F = 0.11 \pm 0.01 $ {\AA}$^{-1}$. 
The small difference in Fermi vectors between $\Gamma$ and Z suggests a modest 
$k_z$ dispersion in undoped BaFe$_2$As$_2$ a point which we will return to later. 
\begin{figure}
	\centering
		\includegraphics[width=0.45\textwidth]{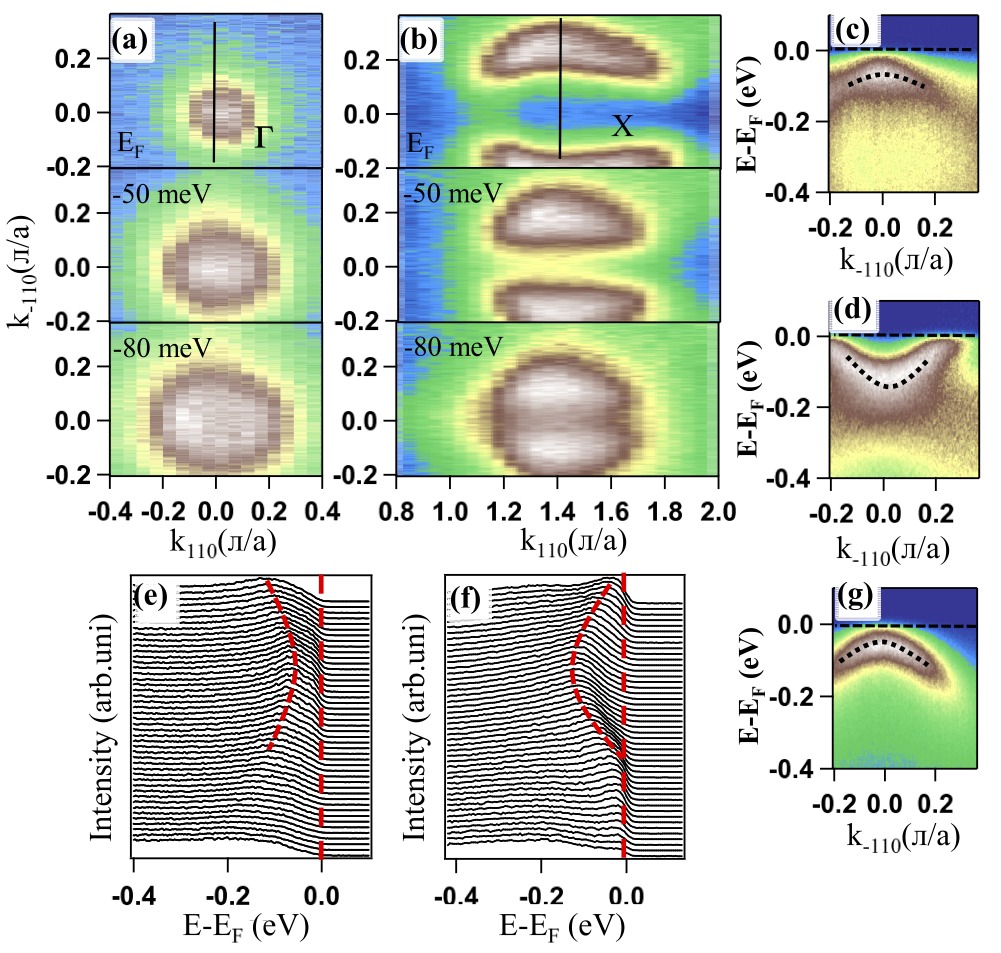}
			\caption{(color online) BaFe$_{1.6}$Co$_{0.4}$As$_2$ARPES data along $\Gamma$-X. (a) and (b) constant energy 
			maps near $E_F$, at $E_F$-50 meV, and $E_F$-80 meV around $\Gamma$ and X, measured with p-polarized and 
			s-polarized photons, respectively. (c) and (d) depict cuts taken along the $k_{-110}$ 
			direction from the center of $\Gamma$ and X while (e) and (f) show the corresponding energy distribution curves. 
			(g) as in panel (c) but measured with s-polarized photons.}
	\label{fig:BCFA}
\end{figure}

Next we discuss our ARPES measurements on BaFe$_{1.6}$Co$_{0.4}$As$_2$. The data shown in Fig. 3 were 
recorded along $\Gamma$-X with h$\nu$=75 eV, corresponding 
to $k_z\approx$0. Constant energy contours around $\Gamma$ were taken with p-polarized photons 
[see Fig.~3(a)] and
analogous contours around X were taken with s-polarized photons [Fig.~3(b)] because no spectral weight 
was observed at X with p-polarized photons. 
\begin{figure*}[th]
	\centering
		\includegraphics[width=0.95\textwidth]{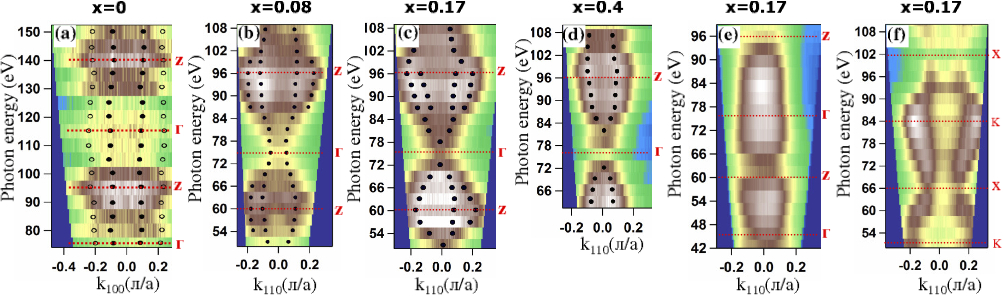}
			\caption{(color online) Photon energy dependent ARPES measurements performed on BaFe$_{2-x}$Co$_x$As$_2$ to reveal the $k_z$ dispersion as a function of doping concentration. (a) Fermi surface for $x$=0, in the $k_{100}$ vs. h$\nu$ plane. (b), (c), and (d) Fermi surfaces in the $k_{110}$ vs. h$\nu$ plane, measured with p-polarized photons, for $x$=0.08, 0.17, and 0.4, respectively. (e) and (f) Fermi surface maps for $x$=0.17 measured with s-polarized photons around $k_x=k_y=0$ and $k_x=k_y=1$ respectively.}
	\label{fig:kz-disp}
\end{figure*}
To resolve the band features 
we made different cuts along  the $k_{-110}$ direction at $\Gamma$ and X. In Fig.~3(c) and (e) we
can see that no bands cross $E_F$ near $\Gamma$ in these measurements performed with p-polarized
photons. The final spectral weight observed in Fig.~3(a) at $E_F$ is due to the tail of the top of the valence band. Using the polarization dependent selection rules of Table I, we can thus 
exclude hole pockets formed from
states having ${z^2}$, ${xz}$, or ${yz}$ character. The existence of hole pockets
with ${x^2-y^2}$ character, which according to the band structure calculations should be
the first to sink below $E_F$ upon electron doping, is excluded from measurements using s-polarized 
photons [Fig.~3(g)]. Therefore for 
high Co doping of BaFe$_2$As$_2$, the hole pockets for $k_z\approx$0 are completely filled, i.e., there are 
no states at $E_F$ near the $\Gamma$ point.  Similar conclusions were derived 
in a previous report \cite{Sekiba} but without polarization analysis, the absence of a 
hole pocket at $\Gamma$ cannot definitely be excluded.
Furthermore, in BaFe$_{1.6}$Co$_{0.4}$As$_2$ the size of the electron 
pocket at X ($k_z\approx$0) has become larger by a factor of two when compared to the undoped system. 
This indicates a shift of
$E_F$ to higher energies upon substituting Fe by Co. On the other hand, we point out
that we were not able to perform a full integration of the volume of the Fermi cylinders
 to judge whether a Co atom really adds one full electron to the Fe 3$d$ dominated low-energy band structure.

In order to reveal the $k_z$ dispersion of the bands in BaFe$_2$As$_2$, we have 
performed photon energy dependent scans with excitation energies ranging from 75 eV to 150 eV 
in steps of 5 eV. Figure~4(a) depicts the Fermi surface map in the $k_{100}$ vs. h$\nu$ 
plane. The data were recorded along the $\Gamma$-M direction with p-polarized light. 
The strong and periodic intensity 
variation with $k_z$ would at first sight seem to indicate a strong dispersion
along this $k$ direction. However, by fitting the MDCs we traced two bands across $\Gamma$ 
and Z which are showing almost no or only minor $k_z$ 
dispersion within our resolution limits. The clue to what is happening here comes from the DFT calculations.
Inspection of Fig.~1(a) shows that near the Z point there is a ${z^2}$ related Fermi surface, while 
near $\Gamma$, only $x^2-y^2$ and $xz/yz$ states cross $E_F$. This radical $k_z$ dependent
change of the orbital character of the Fermi surface -- which is naturally periodic in the c-axis reciprocal
lattice vector -- combined with the strong photon energy dependence of the matrix elements for emission 
from $z^2$ states~\cite{Fink} gives rise to the strong intensity variations shown here and reported 
in \cite{CLiu:2009,Vilmercati:2009,Kondo:2009}. The true $k_z$ dispersion for undoped BaFe$_2$As$_2$ between $\Gamma$
and Z is small, as can also be seen from the DFT results of Fig.~1(c). Analogous ARPES data for SrFe$_2$As$_2$ 
and EuFe$_2$As$_2$ (not shown) confirm that the $k_z$ dispersion derived from a careful MDC
fit analysis for the $k_x=k_y=0$ centered Fermi surfaces is small. Thus the picture
for undoped BaFe$_2$As$_2$ is clear: there is  little $k_z$ dispersion of the states around $k_x=k_y=0$, 
but there is an important $k_z$-dependent change in the orbital character of the $\Gamma$/Z-centerd 
Fermi surfaces.

Having dealt with the important question of the dimensionality and the orbital character of the central
Fermi surface cylinders of the undoped parent compound BaFe$_2$As$_2$, we now address the issue of the 
evolution of the $k_z$ dispersion in BaFe$_{2-x}$Co$_x$As$_2$ as a 
function of Co doping concentration. Figures~4(b)-(d) show the Fermi surface maps 
in the $k_{110}$ vs. h$\nu$ plane, measured
along $\Gamma$-X using p-polarized photons and 
photon energy steps of 3 eV. For lower Co concentrations, by fitting the MDCs, we could resolve  
two bands crossing $E_F$ around Z while at $x$=0.4 only one band could be resolved.
A remarkable observation is that with increasing Co concentration, the $k_z$ dispersion increases
and the spectral weight at $\Gamma$ decreases. In the geometry used in Fig.~4(b)-(d), we probe
${xz}$, ${yz}$, and  ${z^2}$ states. Thus we relate the outer bands
to ${z^2}$ states while the inner ones to ${xz}$, ${yz}$ states.
In order to obtain more information on the orbital character of the bands,
we contrast the $x$=0.17 data of Fig.~4(c), measured with p-polarization, with analogous data
measured with s-polarization which are presented in Fig.~4(e). In the latter the spread of 
the spectral weight along the $k_{110}$ direction is considerably reduced 
and the intensity is large at
$\Gamma$ and small at Z. Since in $\Gamma$-X (s-polarized) geometry, we do not 
detect ${z^2}$ states, and since at that doping level
the ${x^2-y^2}$ band is already completely filled [see also Fig.~1(b)], we observe 
here only the degenerate ${xz}$, ${yz}$ bands. These states give rise to spectral 
weight at $\Gamma$ but not at Z, as we have learned that these states transform at Z into 
${z^2}$ states which cannot be detected in this geometry. Thus summarizing the situation for the states 
near the $\Gamma$ point: with increasing doping concentration,
first the ${x^2-y^2}$ hole pocket will be filled at $x\approx$0.1 and later on the
${xz}$, ${yz}$ hole pocket moves below $E_F$ above $x\approx$0.2. At Z,
the ${x^2-y^2}$ pocket disappears near $x$=0.2 but there remains a hole pocket which 
has predominantly ${z^2}$ character. This means that with
increasing doping concentration the system transforms from a more two dimensional
system with strong nesting conditions to a more three-dimensional metal where nesting is in principle 
possible in the $k_z=\pm$1 planes of the BZ, but there it is strongly reduced due to the different orbital
character of the Fermi surfaces. In 
the $k_z$=0 plane nesting is no longer possible since there is no hole pocket. 
The observed doping dependence of the electronic structure is in remarkable agreement with the
band structure calculations [Fig.~1 (c) and (d)]. Finally, we report for $x$=0.17 also a clear $k_z$ dispersion 
along the X-K direction, derived from fits of MDCs of
data shown in Fig.~4(f) where, due to this specific geometry, we probe all 
states (${xz}$, ${yz}$, and  ${x^2-y^2}$) expected at 
$E_F$. The strength of 
the dispersion and the small difference in the dispersion for the three bands is in good qualitative
agreement with the calculations presented in Fig.~1(f).

In conclusion, we have performed a systematic photon energy dependent high resolution ARPES study 
to reveal the intrinsic $k_z$ dispersion in BaFe$_{2-x}$Co$_x$As$_2$  compounds. 
In the undoped system we see a modest $k_z$ dispersion near $\Gamma$ indicating a more 
two-dimensional system. At higher Co doping the $k_z$ dispersion increases and a gradual 
filling of all three hole pockets at $\Gamma$ is detected. Thus in the region of the BZ near $k_z$=0
the nesting conditions are strongly reduced. In this context, our results
on the orbital character indicate that nesting occurs predominantly in the $k_z$=0 plane  
since at the top and at the bottom of the BZ the orbital character of the Fermi 
cylinders in the center and in 
the corner is different (${x^2-y^2}$/ ${z^2}$ at Z
and ${xz}$/ ${yz}$ at K). Furthermore, the warped Fermi cylinders in optimally doped 
BaFe$_{2-x}$Co$_x$As$_2$ could lead to nesting vectors which are pointing out of the Fe 
layers possibly leading to a more three-dimensional superconducting state.
The present results could 
possibly lead to a microscopic understanding of the disappearance of AFM order 
upon doping for $x$=0.06 and the suppression of the superconducting phase in the overdoped 
phase near $x$=0.18 \cite{Chu}. 
The results on the dimensionality have also implication for the potential application 
of these materials.

\end{document}